\pdfoutput=1

\documentclass[11pt]{article}

\usepackage[final]{acl}

\usepackage{times}
\usepackage{latexsym}

\usepackage[T1]{fontenc}

\usepackage[utf8]{inputenc}

\usepackage{microtype}
\usepackage[inline]{enumitem}
\usepackage{graphicx}
\usepackage{ulem}

\title{\textit{Human-Centric NLP or AI-Centric Illusion?}: \\ A Critical Investigation}

\author{Piyapath T Spencer\\
  Language and Information Technology Programme, Faculty of Arts, CU, Thailand \\
  Center for Information and Language Processing (CIS), LMU Munich, Germany \\
  \texttt{linguistics@piyapath.uk}
}
\date{}

\begin{document}
\maketitle

\begin{abstract}
Human-Centric NLP often claims to prioritise human needs and values, yet many implementations reveal an underlying AI-centric focus. Through an analysis of case studies in language modelling, behavioural testing, and multi-modal alignment, this study identifies a significant gap between the ideas of human-centricity and actual practices. Key issues include misalignment with human-centred design principles, the reduction of human factors to mere benchmarks, and insufficient consideration of real-world impacts. The discussion explores whether Human-Centric NLP embodies true human-centred design, emphasising the need for interdisciplinary collaboration and ethical considerations. The paper advocates for a redefinition of Human-Centric NLP, urging a broader focus on real-world utility and societal implications to ensure that language technologies genuinely serve and empower users.
\end{abstract}

\section{Introduction}
``Human-Centric NLP'' purportedly aims to develop language technologies that are more aligned to human needs, cognition, and behaviour \cite{hovy-spruit-2016-social}. Some argue that by incorporating human factors into NLP systems, we can create more effective, ethical, and user-friendly language technologies \cite{jurgens-etal-2019-just, IntroHumanNLP, kotnis2022, wang-etal-2021-putting}. However, a critical examination of current practices and research trends in this field raises a provocative question: \textit{Is Human-Centric NLP truly centred on human needs, or is it a mere AI-centric illusion?}

Consider, for instance, the development of LLMs like GPT-4 \cite{gpt4}. Even though it appeared as a step towards more human-like language understanding through various tasks, these models primarily focus on improving performance metrics such as perplexity and accuracy on benchmark tasks. The human element often comes into play only in the form of massive datasets used for training or in post-hoc attempts to align the model with human preferences. This approach, whilst impressive in results, arguably prioritises AI capabilities over addressing fundamental human communication needs or cognitive processes.

This paper argues that despite its name and stated intentions, much of what is labeled as Human-Centric NLP is, in fact, predominantly AI-centric. Rather than genuinely prioritising human needs and experiences, these approaches often incorporate human information primarily as a means to enhance AI performance. This misalignment between the proclaimed human-centric goals and the AI-centric reality has significant implications for the development, application, and societal impact of NLP technologies.

Through a critical investigation of current research trends, methodologies, and case studies, the paper aims to address the AI-centric nature under the surface of human-centricity in NLP. The paper also examines how human data and behaviour are often exploited to improve NLP systems without necessarily addressing core human needs or concerns, as  observed by \citet{bender-koller-2020-climbing} and \citet{parrots}. Furthermore, The paper explores the ethical implications of this mischaracterisation and propose a framework for what ``truly'' human-centric NLP might entail, building on the work of scholars who have called for more genuine engagement with human factors in AI development \cite{Crawford2016}. Ultimately, this paper seeks to stimulate a re-evaluation of priorities in NLP research and development. It calls for a genuine shift towards human-centric approaches that place human needs, experiences, and well-being at the forefront, rather than treating them as mere tools for technological advancement.

\section{The Promise of Human-Centric NLP}
The concept of Human-Centric NLP emerged as researchers recognised the need to align language technologies more closely with human needs and cognitive processes, partly as a response to criticisms that they were too focused on technical performance metrics at the expense of real-world applicability and human factors. \citet{kotnis2022} related NLP with the idea of Human-Centric Research (HCR) with the objective to ``place \textit{all} [human] stakeholders at the centre of research''. This paradigm shift promised to conduct research (and create technologies) that are more intuitive, ethical, and aligned with human cognitive processes and societal needs.

Ever since, advocates of Human-Centric NLP have made strong claims about its potential benefits. For instance, \citet{sap-etal-2020-commonsense} and \citet{Kaushik2023} argued that incorporating human knowledge and reasoning patterns could lead to more robust and generalisable NLP systems. They suggested that such systems would be better equipped to handle the nuances and contextual complexities of human language. Moreover, Human-Centric NLP has been acted as a solution to ethical concerns in AI development. \citet{hovy-yang-2021-importance} proposed that by centring human values and societal impact in the design process, we could create more responsible and fair language technologies. 

The promise of Human-Centric NLP extends beyond improved performance. Researchers have argued that this approach could lead to more ethical and socially responsible AI systems. For example, \citet{parrots} in their influential paper on the dangers of large language models, emphasised the need for NLP research to centre on human values and societal impact.

As the field of Human-Centric NLP continues to evolve, researchers are exploring ways to balance technical advancements with ethical considerations and user-centred design. This approach represents a shift from purely performance-driven metrics to a more holistic view of NLP's role in society. There is, nevertheless, ongoing work to translate these human-centric ideals into practical implementations across various NLP applications.

\section{The Reality: AI-Centricity in Disguise}
Whilst the concept of Human-Centric NLP promises an optimistic picture of language technologies aligned with human needs and values, a closer examination of current practices reveals a different reality. Despite the rhetoric of human-centricity, many NLP systems and research directions continue to prioritise AI performance over genuine human-centric considerations.

The development and deployment of such large language models (LLMs) as GPT-series, from GPT-3 \cite{brown2020languagemodelsfewshotlearners} to GPT-4 \cite{gpt4}, serve as a prime example of this disconnect. These models have achieved impressive results on various NLP tasks, yet their development process and application raise serious questions about their alignment with human-centric principles. The data collection methods for these models often involve web scraping vast amounts of information without adequate consideration for privacy, consent, or representation issues. Once again, \citet{parrots} argue that this approach to data collection reflects a prioritisation of model performance over ethical considerations and diverse human perspectives.

Furthermore, the evaluation of these models primarily relies on performance metrics for benchmark tasks and leaderboards. As \citet{ethayarajh-jurafsky-2020-utility} point out, these metrics often fail to capture real-world utility or alignment with human values, instead focusing on narrow technical capabilities. The emphasis on benchmark performance over real-world applicability as such shows the AI-centric nature of current NLP practices.

The resource allocation for training LLMs also reflects a focus on pushing the boundaries of AI capabilities rather than addressing specific human needs or environmental concerns. \citet{strubell-etal-2019-energy} highlight the immense computational resources required for training  models, raising questions about the prioritisation of AI advancement over other important human-related considerations such as environmental impact or more targeted, human-centric applications of NLP technologies.

Beyond LLMs, the field of sentiment analysis provides another example of this disconnect. Tools developed for understanding human emotions often reduce complex affective states to simplistic binary (positive/negative) classifications to ease the computation instead of capture intricate human emotional experiences, making this a reductionist approach that reflects a preference for computational efficiency over truly capturing the complexity of human sentiment.

This gap between the stated goals of Human-Centric NLP and its practical implementation raises critical questions about the field's direction. Are we truly developing technologies that serve human needs, or are we simply creating more sophisticated AI systems that give the illusion of human-centricity? Critics like \citet{BIRHANE2021100205} argue that the focus on technical advancements often overshadows crucial discussions about the societal implications of these technologies.

It becomes more clear that bridging this gap between the promise and reality of Human-Centric NLP requires a fundamental re-evaluation of  priorities and practices in the field. The challenge lies in aligning the impressive technical capabilities of modern NLP systems with genuine human-centric principles that prioritise ethical considerations, user needs, and societal impact.

\section{Some Cases to be Discussed}
To further substantiate the critical examination of Human-Centric NLP, this section presents three cases that exemplify the correlations between human-centric pictures and AI-centric realities.

\noindent \textbf{On Linguistic Varieties} The first case study is \citet{Italian}, addressing the challenges of developing NLP technologies for the diverse language varieties in Italy. Although the author makes important arguments about the technological challenges of Italy's linguistic diversity, his focus on NLP solutions overlooks crucial socioeconomic factors that influence language vitality and also pays insufficient attention to intergenerational transmission dynamics and the predominantly oral dialects, potentially marginalising these aspects in favour of written forms that are more amenable to current NLP techniques. Lacking of a clear framework for community-driven priorities raises questions about how speaker communities themselves might shape research agendas and tool development. Perhaps most tellingly, the approach, whilst interdisciplinary in intent, does not fully integrate insights from sociolinguistics, anthropology, and cultural studies – disciplines crucial for understanding the human dimensions of language use and preservation. Despite its aims, the study remains primarily anchored in a technology-first paradigm that may not fully capture or address the complex human realities of Italy's linguistic landscape.

\noindent \textbf{On Evaluation} The second case study is \citet{ribeiro-etal-2020-beyond}. This paper introduces CheckList, a task-agnostic methodology for testing NLP models; whilst innovative in its approach to NLP model evaluation, reveals several limitations in its human-centricity. The automated testing and model failures, as well as the predefined linguistic capabilities and test types, may not fully capture the nuanced, contextual nature of human language use and potentially oversimplify the complex, holistic nature of human communication. CheckList's black-box testing approach which focusing on discrete linguistic phenomena risks perpetuating a disconnect between model development and the lived experiences of language users. The benchmark-centric view, contrasting differences between model performance and human-like understanding, doesn't deeply explore how these issues relate to real-world language use. Furthermore, the user studies primarily focus on CheckList's ability to generate more tests and uncover bugs, rather than on how it improves the user experience or addresses human-centric language needs.

\noindent \textbf{On Human Data} Our third case study is \citet{takmaz-etal-2020-generating}, aiming to improve image captioning by incorporating human gaze data, ostensibly making the process more `human-centric'. The authors use eye-tracking data to guide the image captioning model, arguing that this approach better aligns with human attention patterns. Its heavy reliance on eye-tracking data as a proxy for human attention risks oversimplifying the complex cognitive processes involved in image interpretation and description. Although the study introduces sequential processing of gaze patterns, this approach also potentially oversimplifies the non-linear and iterative nature of human thought processes during image description tasks. Besides, the introduction of the SSD metric further demonstrates a focus on quantifiable outcomes rather than qualitative alignment with human linguistic behaviour. Notably, the paper gives limited consideration to individual differences such as cultural background, personal experiences, or emotional responses that significantly influence image interpretation. The paper apparently emphasises on improving AI performance through gaze data suggesting a prioritisation of technological advancement over a deeper understanding of human cognition. Apart from this, there features no  discussion on real-world applications for this particular innovation, seemingly the AI-centric nature of the approach. 

These case studies collectively demonstrate the ongoing challenges in achieving Human-Centric NLP. They suggest that \textit{true} human-centricity requires more than just improved performance metrics or the incorporation of human data. Instead, it demands a deep engagement with the complexities of human cognition, cultural contexts, and social dynamics. 

\section{Rethinking \textit{Human-Centricity}}
As critically examining the concept and implementation of Human-Centric NLP, several key questions emerge that need further discussion. These questions challenge the understanding of what it means for NLP to be truly human-centric and how it relates to broader concepts of human-centred design and real-world impact.

\noindent \textbf{1. Is Human-Centric NLP Human-Centred Design?} \\
Human-Centred Design (HCD) is an approach that puts human needs, capabilities, and behaviours at the forefront of the design process. Whilst Human-Centric NLP claims to prioritise human factors, it's debatable whether current practices truly align with HCD principles.

Traditional HCD involves extensive user research, iterative prototyping, and continuous user feedback \cite{HCD}. However, much of Human-Centric NLP research focuses on improving model performance on human-generated datasets or incorporating human-like features, rather than directly involving users in the design process. The case study on Italian language varieties \cite{Italian} demonstrates this disconnect: whilst aiming to address human linguistic diversity, the approach remains largely technology-driven rather than user-driven.

To truly embody HCD, Human-Centric NLP might need to shift towards more participatory research methods, involving end-users throughout the development process, from problem definition to solution evaluation.

\noindent \textbf{2. Does Human-Centric NLP use Human as Another Metrics/Benchmark?} \\
There is a growing concern that Human-Centric NLP often reduces human factors to another set of metrics or benchmarks, rather than genuinely centring human needs and experiences. The CheckList methodology \cite{ribeiro-etal-2020-beyond} exemplifies this tension: it aims to test NLP models on human-like language tasks; however, it still fundamentally treats human language abilities as a one of the benchmark for AI performance.

Similarly, the study on gaze-guided image captioning \cite{takmaz-etal-2020-generating} uses human eye-tracking data to improve AI performance, but it is questionable whether this truly captures the essence of human image interpretation or merely uses human behaviour as another optimisation target.

This trend risks oversimplifying the complexity of human language and cognition. A more genuinely human-centric approach might involve developing evaluation methods that go beyond performance metrics to assess the real-world utility and social impact of NLP systems.

\noindent \textbf{3. Should Human-Centric NLP Take a Step Out from the Computer/Virtual World?} \\
Human-Centric NLP often focuses on improving language technologies within digital environments. However, language is fundamentally a tool for human interaction in the physical world. There is indeed a pressing need for Human-Centric NLP to consider its impacts and applications beyond the virtual realm. \citet{Italian} touches on this by addressing real-world linguistic diversity, but there is potential to go further notwithstanding. 

This could involve studying the real-world consequences of NLP systems, such as their impact on human communication patterns or social dynamics. Developing NLP applications that bridge the digital and physical worlds, like improved assistive technologies for individuals with disabilities,  considering the environmental and societal impacts of large-scale NLP models and infrastructures.\\

These discussions point upon the need for a fundamental re-evaluation of what constitutes Human-Centric NLP. Moving forward, the field should strive for a more holistic approach that truly embodies human-centred design principles, goes beyond using humans as mere benchmarks, and actively engages with the physical world implications of language technologies.

\section{Conclusion}
The paper have examined the disconnect between the promise of Human-Centric NLP and its current implementation. The analysis reveals that many so-called human-centric approaches in NLP remain fundamentally AI-centric nowadays. The case studies and subsequent discussion have demonstrated several key issues: the misalignment with true human-centred design principles, the reduction of human factors to mere benchmarks, and the limited consideration of real-world, physical impacts of NLP technologies. These findings indicate the need for a fundamental reframing of Human-Centric NLP. To address these challenges, we propose that truly Human-Centric NLP should embrace genuine human-centred design methodologies, develop holistic evaluation frameworks, expand its scope to consider broader societal implications, prioritise interdisciplinary collaboration, and centre ethical considerations throughout the development process.  Only then can we hope to develop NLP systems that genuinely serve and empower humans in their diverse contexts and fulfil the true promise of human-centricity in NLP.

\bibliography{reference}

\end{document}